\documentclass[prl,aps,showpacs,twocolumn,superscriptaddress,10pt]{revtex4-1}
\usepackage{amsmath}
\usepackage{epsfig}
\usepackage{bbold}
\begin{document}

\title{Displacement operator for quantum systems with position-dependent mass}

\author{R. N. Costa Filho} \email{rai@fisica.ufc.br}
\affiliation{Departamento de F\'isica, Universidade Federal do
  Cear\'a, Caixa Postal 6030, Campus do Pici, 60455-760 Fortaleza,
  Cear\'a, Brazil}
\author{M. P. Almeida} \email{murilo@fisica.ufc.br}
\affiliation{Departamento de F\'isica, Universidade Federal do
  Cear\'a, Caixa Postal 6030, Campus do Pici, 60455-760 Fortaleza,
  Cear\'a, Brazil}
\author{G. A. Farias} \email{gil@fisica.ufc.br}
\affiliation{Departamento de F\'isica, Universidade Federal do
  Cear\'a, Caixa Postal 6030, Campus do Pici, 60455-760 Fortaleza,
  Cear\'a, Brazil}
\author{J. S. Andrade Jr.}  \email{soares@fisica.ufc.br}
\affiliation{Departamento de F\'isica, Universidade Federal do
  Cear\'a, Caixa Postal 6030, Campus do Pici, 60455-760 Fortaleza,
  Cear\'a, Brazil}

\begin{abstract}
  A translation operator is introduced to describe the quantum
  dynamics of a position-dependent mass particle in a null or constant
  potential. From this operator, we obtain a generalized form of the
  momentum operator as well as a unique commutation relation for $\hat
  x$ and $\hat p_\gamma$. Such a formalism naturally leads to a
  Schr\"odinger-like equation that is reminiscent of wave equations
  typically used to model electrons with position-dependent
  (effective) masses propagating through abrupt interfaces in
  semiconductor heterostructures. The distinctive features of our
  approach is demonstrated through analytical solutions calculated for
  particles under null and constant potentials like infinite wells in
  one and two dimensions and potential barriers.
\end{abstract}

\maketitle

The concept of non-commutative coordinates as a way of removing
divergences in field theories through a universal invariant length
parameter was originally proposed by Heisenberg \cite{Heisenberg}.
This idea led to space-time quantization \cite{Snyder}, where a
non-commutative space operator is defined to allow for the development
of a theory invariant under Lorentz transformation, but not invariant
under translations. Accordingly, the last property leads to continuous
space-time coordinates. Since then, the need for a fundamental length
scale became evident in different areas of physics, such as relativity
\cite{Camelia}, string theory \cite{Witten}, and quantum gravity
\cite{Garay}. In the particular case of quantum mechanics, this
minimum length scale yields a modification in the position momentum
commutation relationship \cite{Kempf,Hinrichsen,Kempf1,Chang}.

Previous studies have shown that the modification of canonical
commutation relations or any modification in the underlying space
typically results in a Schr\"odinger equation with a
position-dependent mass \cite{Quesne}. This approach has been rather
effective in the description of electronic properties of semiconductor
\cite{Bastard} and quantum dots \cite{Serra}. Under this framework,
mass is turned into an operator that does not commute with the
momentum operator. This fact immediately raises the problem of the
ordering of these operators in the kinetic energy operator \cite{Rai}.
In this Communication, we introduce the concept of non-additive
spatial displacement in the Hilbert space. This property not only
changes the commutation relation for position and momentum, which
leads to a modified uncertainty relation, but also reveals a
Schr\"odinger-like differential equation that can be interpreted in
terms of a particle with position-dependent mass, leading to a natural
derivation of the kinetic operator for such problems.

Consider a well-localized state around $x$ that can be changed to
another well-localized state around $x+a+\gamma ax$ with all the other
physical properties unchanged, where the parameter $\gamma$ is the
inverse of a characteristic length that determines the mixing between
the displacement and the original position state. For $\gamma\neq 0$,
the displacement depends explicitly on the position of the system,
while $\gamma=0$ corresponds to a standard translation. This process
can be mathematically expressed in terms of the operator
$\mathcal{T}_\gamma(a)$ as,
\begin{equation}
\mathcal{T}_\gamma (a)|x\rangle=|x+a+\gamma ax\rangle.
\end{equation}
The composition of displacements through $\mathcal{T}_\gamma$ in terms
of two successive infinitesimal translations results in
\begin{equation}
\label{operator1}
\mathcal{T}_\gamma (dx')\mathcal{T}_\gamma (dx'')
=\mathcal{T}_\gamma (dx'+dx''+\gamma dx'dx''),
\end{equation}
which clearly shows the non-additivity characteristic of the operator.
It is also important to note that the inverse operator is given by,
\begin{equation}
\mathcal{T}^{-1}_\gamma (dx)|x\rangle=\left\lvert\frac{x-dx}
{1+\gamma dx}\right\rangle.
\end{equation}
Another relevant property is that $\mathcal{T}_\gamma$ becomes an
identity operator when the infinitesimal translation goes to zero,
\begin{equation}
\lim_{dx\rightarrow 0}\mathcal{T}_\gamma (dx)=\mathbb{1}.
\end{equation}
At this point, we observe that the operator $\mathcal{T}_{\gamma}(x)$
is the infinitesimal generator of the group represented by the
so-called $q$-exponential function originally defined as
\cite{Borges},
\begin{equation}
\label{qexp}
\exp_{q}(x)\equiv\left[1+(1-q)x\right]^\frac{1}{1-q},
\end{equation}
where $\exp_{q}(a)\exp_{q}(b)=\exp_{q}[a+b+(1-q)ab]$, and
$\exp_{0}(a)=\exp(a)$. The definition~(\ref{qexp}) represents a
crucial ingredient in the mathematical formalism of the generalized
Tsallis thermostatistics \cite{Tsallis0} and its several applications
related with non-additive physical systems
\cite{Tsallis1,Tsallis2,Hasegawa}. Associating $\mathcal{T}_\gamma
(dx)$ with the $q$-exponential, and expanding it to first order in
$dx$ leads to
\begin{equation}
\label{infinitesimal}
\mathcal{T}_\gamma (dx)\equiv \mathbb{1}-\frac{i\hat{p}_\gamma dx}{\hbar},
\end{equation}
where $\hat{p}_\gamma$ is a generalized momentum operator and we are
using the fact that momentum is a generator of translation. Now,
considering that
\begin{equation}\hat{x}\mathcal{T}_\gamma (dx)|x\rangle=
(x+dx+\gamma xdx)|x+dx+\gamma xdx\rangle,
\end{equation}
and $\mathcal{T}_\gamma (dx)\hat{x}|x\rangle =x|x+dx+\gamma
xdx\rangle$, we obtain the commutation relation,
\begin{equation}
\left[\hat{x},\mathcal{T}_\gamma (dx)\right]|x\rangle\simeq
dx(1+\gamma x)|x\rangle,
\end{equation}
where the error is in second order in $dx$. Then we get,
\begin{equation}
\left[\hat{x},\hat{p}_\gamma\right]= i\hbar(1+\gamma x),
\end{equation}
with the following uncertainty relation,
\begin{equation}
\Delta x\Delta p_\gamma\ge \frac{\hbar}{2}\left(1+\gamma \langle x\rangle
\right),
\end{equation}
which expresses the fact that the uncertainty in a measurement depends
on $\gamma$ as well as on the average position of the particle.

Next, it is straightforward to obtain an expression for the modified
momentum operator in the $x$-basis using the definition
(\ref{infinitesimal}),
\begin{equation}
\label{momentum}
 \hat{p}_\gamma |\alpha\rangle=-i\hbar(1+\gamma x)\frac{d}{dx}|\alpha\rangle.
\end{equation}
From (\ref{momentum}) the modified momentum operator can be written
for short as $\hat{p}_{\gamma}=-i\hbar D_{\gamma}$ with
\begin{equation}
D_{\gamma}\equiv(1+\gamma x) \frac{d}{dx}
\end{equation}
being a deformed derivative in space. In the $x$-representation, the
corresponding time-dependent Schr\"odinger equation is
\begin{equation}
i\hbar \frac{\partial}{\partial t}\langle x|\alpha,t_s\rangle =\langle x|H|\alpha,t_s\rangle,
\end{equation}
and if we consider the Hamiltonian operator to be,
$H=\hat{p}_\gamma^2/2m+V(x)$, we arrive at the following differential
equation:
\begin{align}
\label{differential}
i\hbar \frac{\partial}{\partial t}\psi(x,t) &=-\left(\frac{\hbar^2}{2m}\right)D_\gamma^2
\psi(x,t)+V(x)\psi(x,t).
\end{align}

We now focus our attention on the case of a single spinless particle
system. If the wave function $\psi(x,t)$ is normalized, it is possible
to define a probability density $\rho(x,t)=|\psi(x,t)|^2$. Using
Eq.(\ref{differential}) it is straightforward to derive a modified
continuity equation
\begin{equation}
\frac{\partial\rho}{\partial t}+D_\gamma J_\gamma=0,
\end{equation}
where the probability flux is given by,
\begin{equation} 
J_\gamma=\frac{\hbar(1+\gamma x)}{2mi}\left(\psi^*\frac{d \psi}{dx}-\psi\frac{d \psi^*}{dx}\right).
\end{equation} 
For standing waves in a null potential, the wave function $\phi(x)$
satisfying (\ref{differential}) obeys
\begin{equation}
-\frac{\hbar^2}{2m}D_\gamma^2\phi(x)=E\phi(x),
\end{equation}
or
\begin{equation}
\label{differential1}
\frac{\hbar^2}{2m_e}\frac{d^2\phi(x)}{dx^2}+\frac{\hbar^2}{2}\frac{d}{dx}\left(\frac{1}{2m_e}\right)\frac{d\phi(x)}{dx}
+E\phi(x)=0,
\end{equation}
with $m_{e}\equiv m/(1+\gamma x)^{2}$ being the particle's {\it
  effective mass}, in perfect analogy with problems involving a
position-dependent mass particle in semiconductor heterostructures
\cite{Bastard}. Here this particular expression for the effective mass
arises naturally from the non-additive translation operator. Equation
(\ref{differential1}) can then be rewritten in the form of the
Cauchy-Euler equation \cite{Boas},
\begin{equation}
\label{cauchy}
u(x)^2\frac{d^2\phi(x)}{du^2}+au(x)\frac{d\phi(x)}{du}+b\phi(x)=0.
\end{equation}
with $u(x)=\left(1+\gamma x\right)$, $a=1$, and
$b=2mE/(\hbar^2\gamma^2)$. The general solution for Eq.(\ref{cauchy})
is
\begin{equation}
\label{pwave}
\phi(x)=\exp{\left[\pm i\frac{k}{\gamma}\ln{(1+\gamma x})\right]},
\end{equation}
where $k$ is a continuous variable regarding the particle's wave
vector. Although the wave function for a free particle is not the
usual plane wave, the particle's energy spectra is continuous,
$E=\frac{\hbar^2k^2}{2m}$, and independent of $\gamma$. For this free
particle, the probability flux $J_\gamma$ is the same as an object
moving at the classical velocity $\hbar k/m$.

If we now assume that the particle is confined to a one-dimensional
infinite well of length $L$, the boundary conditions $\phi(0)=0$ and
$\phi(L)=0$ lead to the wave function,
\begin{equation}
\label{wavewell}
\phi_n(x)=\begin{cases}
A_n\sin\left[\frac{k_n}{\gamma}\ln{(1+\gamma x)}\right], &\quad 0<x<L\\
0, &\quad {\rm otherwise,}
\end{cases}
\end{equation}
where the wave vector is now quantized
\begin{equation}
k_n=\frac{n\pi\gamma}{\ln(1+\gamma L)}, \quad{\rm with}\quad n={1,2,3,4,\dots}
\end{equation}
The energy for a confined particle can then be written as,
\begin{equation}
E_n=\frac{\hbar^2n^2\pi^2\gamma^2}{2m\ln^2(1+\gamma L)},
\end{equation}
where $\ln{(1+\gamma L)}/\gamma$ corresponds to an effective
dilated/contracted well that approaches $L$ as $\gamma \rightarrow 0$.
\begin{figure}[t]
\centering
\includegraphics[scale =0.4]{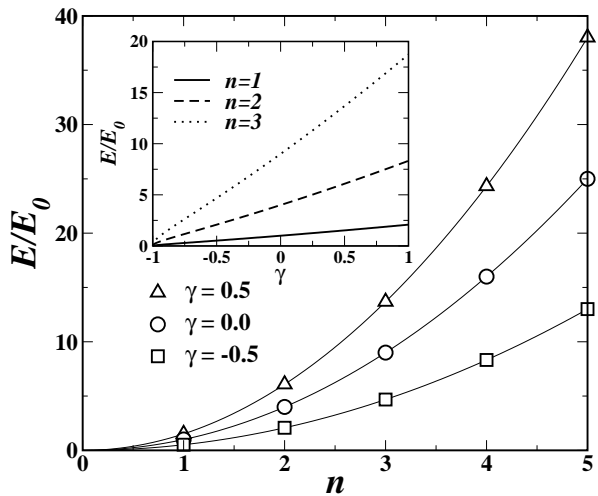}
\caption{The energy for a particle confined in an infinite well. The
  circles represent the energy for $\gamma=0$, the squares for
  $\gamma=-0.5$, and the triangles for $\gamma=0.5$. The energies are
  discrete, the solid lines are just guides to the eye. The inset
  shows the energy against $\gamma$ for the three lowest levels.}
\label{Fig1}
\end{figure}

In Fig.~\ref{Fig1} we show how the energy levels of a particle
confined to an infinite well increase with $n$ for different values of
$\gamma$. Since the effective mass, $m_{e}$, in our description,
decreases with $\gamma$, and the lower the mass, the bigger is the
kinetic energy of the particle, it means that the energy $E_{n}$
increases with $\gamma$, as depicted in the inset of Fig.~\ref{Fig1}.
The same reasoning can be applied to the size of the well, namely, the
increase (decrease) of $\gamma$ above (below) zero, leads to a more
pronounced contraction (dilation) of its effective length. The
asymmetry caused by the parameter $\gamma$ can be adequately
quantified in terms of the average position of the particle in the
box, calculated as $\langle x\rangle=\int_0^L\phi^*x\phi dx$. From
(\ref{wavewell}), we obtain the following expression:
\begin{equation}
  \langle x\rangle=\frac{L}{2}\left[\frac{\gamma^2+4k^2_n}{4(\gamma^2+k^2_n)}-\frac{3}{2}\frac{\gamma}{L(\gamma^2+k^2_n)}\right],
\end{equation}  
and the average of the modified momentum is $\langle
p_\gamma\rangle=0$. Figure 2 shows the average position of the
particle against $\gamma$. As expected, when $\gamma=0$ the average
value is always $0.5$. The ground state is the most affected by the
non-additivity of the space. As the quantum number increases, for
example $n=20$, the particle's average position becomes independent of
$\gamma$, $\langle x\rangle \rightarrow 0.5$.
\begin{figure}[t]
\centering
\includegraphics[scale =0.4]{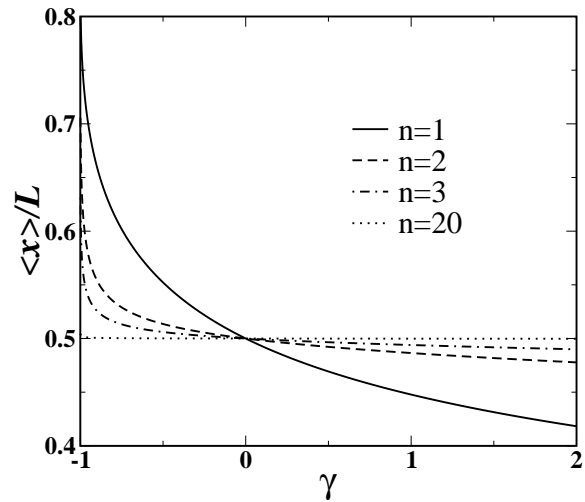}
\caption{The average position $\langle x\rangle$ of a particle
  confined in an infinite quantum well. The solid line gives the
  average value of $x$ for $n=1$, the dashed line is for $n=2$, the
  case $n=3$ is the dashed-dotted line, and the large quantum number
  $n=20$ is represented by the dotted line. As $n$ increases the
  asymmetry of the wave function is reduced, $\langle x\rangle
  \rightarrow 0.5$.}
\label{Fig2}
\end{figure}

\begin{figure}[t]
\centering
\includegraphics[scale=0.32]{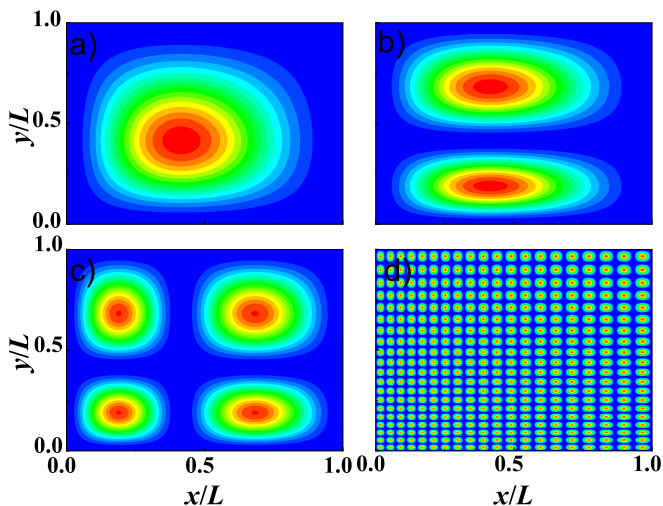}
\caption{The contour plot of the probability density for a particle in
  a two-dimensional box for $\gamma=1$, where the quantum numbers used
  are: a) $n_x=n_y=1$, b) $n_x=1, n_y=2$, c)$n_x=n_y=2$ and d)
  $n_x=n_y=20$. The probability increases from blue to red.}
\label{Fig3}
\end{figure}

The position operator in other space directions still commutes.
Therefore, the theory developed here can be easily extended for two
and three dimensions. For example, when considering a two-dimensional
infinite well, the corresponding wave function can be expressed as the
product $\Phi(x,y)=\phi_1(x)\phi_2(y)$, where $\phi_1(x)$ and
$\phi_2(y)$ are the wave functions in the $x$ and $y$ directions,
respectively. The contour plots for the probability density
$\rho(x,y)$ of a particle moving in a two-dimensional box are shown in
Fig. \ref{Fig3}(a) $n_x=n_y=1$; in fig. \ref{Fig3}(b) for $n_x=1,
n_y=2$ ; in Fig. \ref{Fig3}(c) for $n_x=n_y=2$; and in Fig.
\ref{Fig3}(d) $n_x=n_y=20$, where we have used $\gamma=1$ in all
panels. The ground state shows clearly that the particle spends more
time out of the box center. These results also indicate that the
correspondence principle remains valid, as for $n_x=n_y=20$ (large
quantum numbers) the probability to find a particle is practically the
same everywhere in the square well [see Fig.~\ref{Fig3}(d)].
 
\begin{figure}[t]
\centering
\includegraphics[scale =0.4]{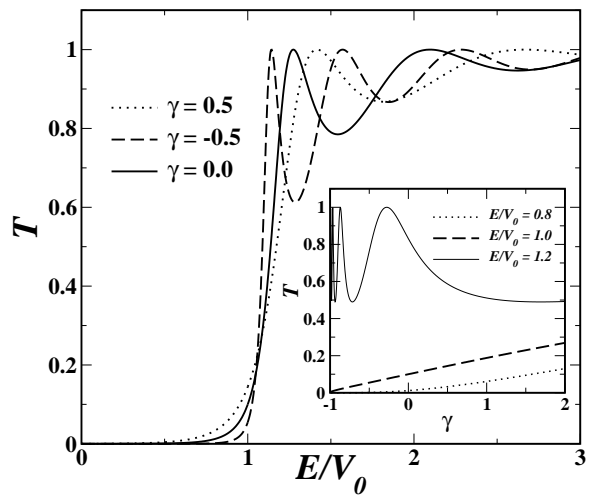}
\caption{The transmission probability $T$ for a rectangular barrier
  for three different values of $\gamma$. The solid line is for the
  normal space ($\gamma=0$). The dotted curve is for $\gamma=0.5$,
  while $\gamma=-0.5$ corresponds to the dashed curve. The inset shows
  the transmission probability for three different values of energy.
  Here we used $\sqrt{2mV_0/\hbar^2}=6 $}.
\label{Transm}
\end{figure}

From the above examples, for a null (free particle) or a constant
potential (infinite 1D and 2D well), the effect of $\gamma$ on the
wave function of the particle is to stretch or contract the space
variable. In this way, we expect the same effect for a particle
subjected to a potential barrier with height $V_0>0$, and located
between $x=0$ and $x=a$. The wave function becomes a linear
combination of Eq.(\ref{pwave}),
\begin{equation}
\phi(x)=\begin{cases}
e^{ika'}+re^{-ika'}, & \mbox{for } x<0\\
Ae^{iqa'}+Be^{-iqa'}, & \mbox{for } 0<x<a\\
te^{ika'}, & \mbox{for } x>a
\end{cases},
\end{equation}
where $k=\sqrt{2mE/\hbar^2}$, $q=\sqrt{2m(E-V_0)/\hbar^2}$, and we
used $a'=\ln{(1+\gamma a)}/\gamma$ for short. The coefficients $A, B,
r, t$ are found taking the continuity of the wave function and its
spatial derivative in $x=0$ and $x=a$. The transmission and tunneling
probability is modified by $\gamma$:
\begin{equation}
T^{-1}=|t|^2=\begin{cases}1+\frac{V_0^2\sin^2{qa'}}{4E(E-V_0)}, & \mbox{for } E>V_0\\
\\
1+\frac{V_0^2\sinh^2{qa'}}{4E(V_0-E)}, & \mbox{for } E<V_0
\end{cases}.
\end{equation}
Figure 4 shows the transmission probability against the energy ratio
$E/V_0$ for different values of $\gamma$. For $E>V_0$, we can see that
the resonances ($T=1$) depend on the value of $\gamma$. Increasing
$\gamma$ is analogous to decrease the length of the barrier potential,
and the transmission probability gets closer to $1$. When $E<V_0$, the
quantum tunneling probability increases with $\gamma$ (thinner
barrier). The inset in Fig. \ref{Transm} shows an oscillation with
increase period for $E>V_0$.

Let's turn our attention to the kinetic operator developed through the
non-additive approach introduced here. According to
Eq.(\ref{momentum}), we can write the modified momentum operator as
$\hat{p}_\gamma=(1+\gamma x)\hat{p}$, so that the kinetic energy
operator becomes,
\begin{equation}
\label{kinetic}
\hat{K}=\frac{1}{2}\frac{1}{\sqrt{m_e}}\hat{p}\frac{1}{\sqrt{m_e}}\hat{p}.
\end{equation}
By comparison, this expression does not constitute a particular case
of the general kinetic energy operator proposed in Ref.~\cite{Roos} to
describe a position-dependent mass in the effective mass theory of
semiconductors, namely,
$\frac{1}{4}\left(m^\alpha\hat{p}m^\beta\hat{p}m^\delta+m^\delta\hat{p}m^\beta\hat{p}m^\alpha\right)$,
with $\alpha+\beta+\delta=-1$. To the best of our knowledge, and
despite its claimed generality in terms of the parameters $\alpha$,
$\beta$, and $\gamma$, this last operator has not been deduced from
first principle calculations.

In summary, we have introduced a non-additive translation operator
that can be identified as a $q$-exponential \cite{Borges}. By means of
this operator, we have developed a modified momentum operator that
naturally leads to a Schr\"odinger-like equation reminiscent of the
wave equation typically used to describe a particle with
position-dependent mass. First, our results indicate that a free
particle in this formalism has a continuum energy spectrum with a wave
function that is a modified plane wave. For a constant potential, like
the problem of a particle confined to an infinite well, the energy now
depends on the parameter $\gamma$, and for potential barrier, the
peaks of maximum transmission probability are $\gamma$-dependent. In
this context, we can argue that the substrate, as defined here,
behaves like a graded crystal whose local properties determine the
effective mass of the confined particle. Our approach can therefore be
useful to describe the particle's behavior within interface regions of
semiconductor heterostructures. Another interesting consequence is
that one can map this non-additive theory to an additive one using the
appropriate potential. As future work, we intend to investigate the
behavior of the non-additive particle system when subjected to
confining potentials that depend explicitly on position, for example,
the harmonic oscillator or the central potential cases.
 
\acknowledgments{We thank Constantino Tsallis for fruitful discussions
  and the Brazilian agencies CNPq, CAPES, and FUNCAP for financial
  support.}

\end{document}